\documentclass[english,onecolumn, smallcondensed, final]{svjour3}
\usepackage[T1]{fontenc}
\usepackage[latin9]{inputenc}
\usepackage{color}
\usepackage{babel}
\usepackage{url}
\usepackage{amsmath}
\usepackage{amssymb}
\usepackage[unicode=true,pdfusetitle,
 bookmarks=true,bookmarksnumbered=false,bookmarksopen=false,
 breaklinks=true,pdfborder={0 0 0},pdfborderstyle={},backref=false,colorlinks=true]
 {hyperref}

\makeatletter
\usepackage[dvipsnames]{xcolor}
\hypersetup{linkcolor=blue,citecolor=blue}
\usepackage{cite}
\usepackage{bm}
\usepackage[final]{microtype}
\usepackage{amsmath}
\usepackage{bm}

\allowdisplaybreaks

\makeatother

\begin{document}

\title{Reflection matrices with $U_{q}[\text{osp}^{\left(2\right)}\left(2|2m\right)]$
symmetry}

\author{R. S. Vieira \and A. Lima-Santos }

\institute{Universidade Federal de São Carlos, Departamento de Física, caixa-postal
676, CEP. 13565-905, São Carlos, SP, Brasil.  \email{R. S. Vieira: rsvieira@df.ufscar.br \and A. Lima-Santos: dals@df.ufscar.br}}
\maketitle
\begin{abstract}
We propose a classification of the reflection $K$-matrices (solutions
of the boundary Yang-Baxter equation) for the $U_{q}[\mathrm{osp}^{\left(2\right)}\left(2|2m\right)]=U_{q}[C^{\left(2\right)}\left(m+1\right)]$
vertex-model. We have found four families of solutions, namely, the
complete solutions, in which no elements of the reflection $K$-matrix
is null, the block-diagonal solutions, the \emph{X}-shape solutions
and the diagonal solutions. We highlight that these diagonal $K$-matrices
also hold for the $U_{q}[\mathrm{osp}^{\left(2\right)}\left(2n+2|2m\right)]=U_{q}[D^{\left(2\right)}\left(n+1,m\right)]$
vertex-model.
\end{abstract}


\tableofcontents{}

\pagebreak{}

\section{Introduction }

The importance of the Yang-Baxter (YB) equation is now a very well
established fact. This equation appeared first in relativistic field
theory as a sufficient condition for the factorization of the scattering
amplitudes for a system of particles interacting via delta potentials
\cite{Mcguire1964,Yang1968,Yang1967,Zamolodchikov1979}. Soon after,
the same equation was derived by Baxter \cite{Baxter1972,Baxter1978}
in the field of statistical mechanics: in this case, the YB equations
ensures that the transfer matrix (a mathematical object related to
the partition function of the model) commutes with itself for different
values of the spectral parameter $x=e^{u}$, so that it can be regarded
as the generator of infinitely many quantities in evolution and the
corresponding model can be regarded as integrable in the sense of
Liouville.

The interest on the YB equation was increased with the formulation
of the quantum inverse scattering method, also known as algebraic
Bethe Ansatz \cite{Faddeev1979,TakhtadzhanFaddeev1979,ZamolodchikovFateev1980,Faddeev1982,Sklyanin1982,Baxter1982,Faddeev1995,Korepin1997,Batchelor2007}.
This powerful technique allows (if applied successfully) the exact
diagonalization of the transfer matrix associated with a given vertex-model,
the YB equation providing the commutation relations between the relevant
operators. The complete diagonalization of the transfer matrix depends,
however, on the solution of the so-called \emph{Bethe Ansatz equations}
$-$ a complex set of non-linear equations whose analytical solution
is not yet available \cite{Vieira2015}. 

More recently, the YB equation proved to be important also in classical
field theory, condensed matter, nuclear physics and in high energy
physics through the \emph{AdS/CFT} correspondence between the $\mathcal{N}=4$
super Yang-Mills gauge theory and the $AdS_{5}\times S^{5}$ sigma
model of string theory \cite{Maldacena1999,MinahanZarembo2003,BeisertStaudacher2003,BenaPolchinskiRoiban2004}.
In pure mathematics, the YB equation contributed to the development
of algebraic structures associated with Lie (super)algebras, for instance
the Hopf algebras and the formulation of quantum groups \cite{Drinfeld1988A,Drinfeld1988B,Faddeev1987,Jimbo1985,Jimbo1986B,LambeRadford2013}.

The YB equation consists in a matrix relation defined on the $\mathrm{End}\left(V\otimes V\otimes V\right)$,
where $V$ is a $N$-dimensional complex vector space, which reads
\cite{Mcguire1964,Yang1967,Yang1968,Baxter1972,Baxter1978,Zamolodchikov1979,GhoshalZamolodchikov1994},
\begin{equation}
R_{12}(x)R_{13}(xy)R_{23}(y)=R_{23}(y)R_{13}(xy)R_{23}(x).\label{PYB1}
\end{equation}
In this equation, $R$ is a matrix defined on $\mathrm{End}\left(V\otimes V\right)$
which is regarded as the solution of the YB equation. The matrices
$R_{12}$, $R_{23}$ and $R_{13}$ are obtained from $R$ through
the expressions $R_{12}=R\otimes I$, $R_{23}=I\otimes R$ and $R_{13}=P_{12}R_{23}P_{12}$,
where $I$ is the identity matrix defined on $\mathrm{End}\left(V\right)$
and $P_{12}=P\otimes I$, with $P$ denoting the permutator matrix
so that $P(A\otimes B)=B\otimes A,\forall\left\{ A,B\right\} \in\mathrm{End}\left(V\right)$.
Solutions of the YB equation have been investigated for a long time
ago $-$see \cite{Baxter1982,Korepin1997,ZamolodchikovFateev1980,KulishSklyanin1980,KulishSklyanin1982,Faddeev1995,Bazhanov1985}
and references therein. Jimbo has already proposed a classification
of the $R$-matrices associated with all non-exceptional affine Lie
algebras \cite{Jimbo1986A}. More recently, supersymmetric solutions
of the YB equation (which are associated with quantum deformations
of affine Lie superalgebras) were also found \cite{Bazhanov1987,BazhanovShadrikov1987}.
In special, Galleas and Martins derived new solutions of the YB equation
that can be regarded as non-trivial graded generalizations of Jimbo's
$R$-matrices \cite{Galleas2004,Galleas2006,Galleas2007}.

The YB equation (\ref{PYB1}) ensures the integrability of a given
vertex-model for periodic boundary conditions. When non-periodic boundary
conditions are present, the integrability of the system at the boundaries
is guaranteed by the \emph{boundary YB equation}, also known as the
\emph{reflection equation}, \cite{Sklyanin1988,Cherednik1984,Mezincescu1991A,Mezincescu1991B},
\begin{equation}
R_{12}\left(x/y\right)K_{1}\left(x\right)R_{21}\left(xy\right)K_{2}\left(y\right)=K_{2}\left(y\right)R_{12}\left(xy\right)K_{1}\left(x\right)R_{21}\left(x/y\right).\label{BYB1}
\end{equation}
This is a matrix equation defined on $\mathrm{End}\left(V\otimes V\right)$
in which $R_{12}$ denotes just the matrix $R$, solution of the periodic
YB equation (\ref{PYB1}), while $R_{21}=PR_{12}P$ and $K_{1}=K\otimes I$
and $K_{2}=I\otimes K$, where the reflection $K$-matrix $-$ the
required solution of the boundary YB equation $-$ is a matrix defined
on $\mathrm{End}\left(V\right)$. The boundary YB equation was introduced
by Sklyanin in \cite{Sklyanin1988} (based on a previous work of Cherednik
\cite{Cherednik1984}) and it was applicable only for symmetric $R$-matrices
$\left(R_{12}=R_{21}\right)$. The boundary YB equation (\ref{BYB1}),
which holds also for the non-symmetric $R$-matrices, was introduced
by Mezincescu and Nepomechie in \cite{Mezincescu1991A}. Solutions
of the boundary YB equation (\ref{BYB1}) have a long history as well.
The first ones were found by Sklyanin himself \cite{Sklyanin1988}
and, since then, the reflection $K$ matrices associated with several
vertex-models were found \cite{deVega1993,deVega1994,Inami1996,Lima1999,Guan2000,Lima2001}.
The quantum group formalism for the non-periodic case is, however,
still in progress \cite{Kulish1992,Nepomechie2002,Doikou2005,Vaes2007}. 

Supersymmetric reflection $K$-matrices were also obtained in the
last two decades, although the graded boundary YB equation \cite{BrackenEtAl1998}
is generally more difficult to be solved. Indeed, although a classification
of the reflection $K$-matrices associated with non-graded affine
Lie algebras has been proposed \cite{Malara2006}, a classification
of the graded reflection $K$-matrices is yet not available. A great
advance towards this end was obtained recently by Lima-Santos in a
series of papers, on which the reflection $K$-matrices of the $U_{q}[\mathrm{sl}^{(2)}\left(r|2m\right)]$,
$U_{q}[\mathrm{osp}^{\left(1\right)}\left(r|2m\right)]$, $U_{q}[\mathrm{spo}\left(2n|2m\right)]$
and $U_{q}[\mathrm{sl}^{\left(1\right)}\left(m|n\right)]$ vertex-models
were derived and classified \cite{Lima2009A,Lima2009B,Lima2009C,Lima2010}. 

The present work can be thought as a continuation of the studies above,
as we present here the reflection $K$-matrices of the supersymmetric
$U_{q}[\mathrm{osp}^{\left(2\right)}\left(2|2m\right)]$ vertex-model.
Since this vertex-model describes a supersymmetric interacting system,
we should take into account the theory of Lie superalgebras \cite{Moody1967,Moody1968,Kac1977A,Kac1977B}
to study its reflection $K$-matrices. In the graded case all the
mathematical operations should be modified accordingly \cite{Bazhanov1987,BazhanovShadrikov1987}.
In a $\mathbb{Z}_{2}$-graded Lie superalgebra, we distinguish even
elements from the odd ones (physically, the even elements describe
bosons, while the odd elements describe fermions). Hence, we decompose
the vector space $V$ as a direct sum of the even and odd part: $V=V_{\mathrm{even}}\oplus V_{\mathrm{odd}}$.
Even and odd elements can be distinguished through the\emph{ Grassmann
parity} defined as, 
\begin{equation}
\pi_{a}=\begin{cases}
1, & a\in V_{\mathrm{odd}},\\
0, & a\in V_{\mathrm{even}}.
\end{cases}
\end{equation}

All matrix operations is redefined in their graded version. For instance,
the graded tensor product of two matrices $A=\in\mathrm{End}\left(V\right)$
and $B=\in\mathrm{End}\left(V\right)$ is defined by 
\begin{equation}
A\otimes_{g}B=\sum_{\left\{ a,b,c,d\right\} =1}^{N}\left(-1\right)^{\pi_{a}\left(\pi_{b}+\pi_{c}\right)}A_{ab}B_{cd}E_{ab}^{cd},
\end{equation}
where $E_{ab}^{cd}=e_{ab}\otimes e_{cd}$ and $e_{ab}$ denotes the
standard Weyl matrix (a matrix whose element on the $a$-th row and
$b$-th column is equal to 1 while the other elements are all zero).
The graded permutator matrix becomes given by 
\begin{equation}
P_{g}=\sum_{\left\{ a,b\right\} =1}^{N}\left(-1\right)^{\pi_{a}\pi_{b}}E_{ab}^{ba}.
\end{equation}
Besides, the graded trace of a matrix $A=\in\mathrm{End}\left(V\right)$
and its graded transposition are defined respectively by 
\begin{equation}
\mathrm{tr}{}_{g}\left(A\right)=\sum_{a}^{N}\left(-1\right)^{\pi_{a}}A_{aa},\qquad A_{g}^{t}=\sum_{\left\{ a,b\right\} =1}^{N}\left(-1\right)^{\left(\pi_{a}+1\right)\pi_{b}}A_{ba}e_{ab}.
\end{equation}

In the graded case, both the periodic as the boundary YB equations
can be written in the same form (\ref{PYB1}) and (\ref{BYB1}), respectively,
if all the linear operations are considered in their graded form \cite{Bazhanov1987,BazhanovShadrikov1987}.
Alternatively, we can introduce the so-called scattering $S$-matrix
through 
\begin{equation}
S(x)=P_{g}R(x),
\end{equation}
so that both the periodic (\ref{PYB1}) as the boundary (\ref{BYB1})
YB equations can be written in a form which is insensitive to the
graduation. In this case, all linear operations should be considered
in their non-graded versions, but the periodic (\ref{PYB1}) and the
boundary YB equation become given respectively by 
\begin{equation}
S_{12}(x)S_{23}(xy)S_{12}(y)=S_{23}(y)S_{12}(xy)S_{23}(x),\label{PYB2}
\end{equation}
and 
\begin{equation}
S_{12}\left(x/y\right)K_{1}\left(x\right)S_{12}\left(xy\right)K_{1}\left(y\right)=K_{1}\left(y\right)S_{12}\left(xy\right)K_{1}\left(x\right)S_{12}\left(x/y\right).\label{BYB2}
\end{equation}

\section{The $U_{q}[\mathrm{osp}^{\left(2\right)}\left(2n+2|2m\right)]=U_{q}[D^{\left(2\right)}(n+1|m)]$
and $U_{q}[\mathrm{osp}^{\left(2\right)}\left(2|2m\right)]=U_{q}[C^{\left(2\right)}(m+1)]$
vertex-models}

In this work we shall consider a $S$-matrix, solution of the periodic
YB equation (\ref{PYB2}), which was obtained by Galleas and Martins
in \cite{Galleas2006}. In that work the authors employed a baxterization
procedure through representations of the dilute Birman-Wenzl-Murakami
algebra \cite{Murakami1987,Birman1989} in order to find new solutions
of the graded YB equation. Among other solutions previously known,
they found a $S$-matrix describing a vertex-model containing $2n+2$
bosons and $2m$ fermions, which was regarded as a supersymmetric
generalization of Jimbo's $S$-matrix \cite{Jimbo1986A}, which is
associated with the $U_{q}[\mathrm{o}^{\left(2\right)}\left(2n+2\right)]=U_{q}[D_{n+1}^{(2)}]$
quantum twisted affine Lie algebra\footnote{The Birman-Wenzl-Murakami algebra was also considered before by Grimm
in \cite{Grimm1993,Grimm1994A,Grimm1994B,Grimm1995}, where other
models related to the $U_{q}[o\left(2n+2\right)^{(2)}]=U_{q}[D_{n+1}^{(2)}]$
symmetry were also obtained.}. 

Employing a simplified notation, the Galleas-Martins $S$-matrix can
be written as follows:
\begin{align}
S_{\left(\kappa_{1},\kappa_{2},\nu\right)}^{m,n}(x) & =\sum_{a,b\,\in\,\sigma_{1}}\left[a_{1}(x)\left(E_{ba}^{ab}+E_{ab}^{ba}\right)+a_{2}(x)\left(E_{ba}^{ab''}+E_{ab}^{b''a}\right)\right]\nonumber \\
 & +\sum_{a,b\,\in\,\sigma_{2}}{\displaystyle a_{3}(x)E_{bb}^{aa}}+\sum_{a,b\,\in\,\sigma_{3}}{\displaystyle a_{4}(x)E_{bb}^{aa}}\nonumber \\
 & +\sum_{a,b\,\in\,\sigma_{4}}\left[a_{5}(x)E_{a''a}^{aa''}+a_{6}(x)E_{aa}^{aa}+a_{7}(x)E_{a''a''}^{aa}+a_{8}(x)E_{aa''}^{aa''}\right]\nonumber \\
 & +\sum_{a,b\,\in\,\sigma_{5}}b_{1}^{a}(x)E_{aa}^{aa}\nonumber \\
 & +\sum_{a,b\,\in\,\sigma_{6}}\left[b_{2}^{a}(x)\left(E_{bb}^{aa}+E_{a''a''}^{b''b''}\right)+b_{3}^{a}(x)\left(E_{b''b}^{aa}+E_{a''a''}^{b''b}\right)\right]\nonumber \\
 & +\sum_{a,b\,\in\,\sigma_{6}}\left[b_{4}^{a}(x)\left(E_{a''b}^{ab''}+E_{ba''}^{b''a}\right)+b_{5}^{a}(x)\left(E_{a''b}^{ab}+E_{ba''}^{ba}\right)\right]\nonumber \\
 & +\sum_{a,b\,\in\,\sigma_{7}}c_{1}^{ab}(x)E_{a''b}^{ab''}+{\displaystyle \sum_{a,b\,\in\,\sigma_{8}}}{\displaystyle c_{2}^{ab}(x)E_{ba}^{ab}},\label{S}
\end{align}
where $\mu=m+n$, $N=2\mu+2$ and we introduced conveniently the notations:
\begin{equation}
a'=N-a+1,\quad b'=N-b+1,\quad a''=N-a+2,\quad b''=N-b+2.
\end{equation}
The sums in the indexes $a$ and $b$ run from $1$ to $N$ and they
are restricted by the subsets $\sigma_{k}$ as defined below: 
\begin{align}
\sigma_{1} & =\left\{ a\neq b,a\neq b'';b=\mu+1\text{ or }b=\mu+2\right\} ,\\
\sigma_{2} & =\left\{ a<b,a\neq b'',a\neq\mu+1,a\neq\mu+2;b\neq\mu+1,b\neq\mu+2\right\} ,\\
\sigma_{3} & =\left\{ a>b,a\neq b'',a\neq\mu+1,a\neq\mu+2;b\neq\mu+1,b\neq\mu+2\right\} ,\\
\sigma_{4} & =\left\{ a=b,a=\mu+1\text{ or }a=\mu+2\right\} ,\\
\sigma_{5} & =\left\{ a=b,a\neq\mu+1,a\neq\mu+2\right\} ,\\
\sigma_{6} & =\left\{ a\neq\mu+1,a\neq\mu+2;b=\mu+1\text{ or }b=\mu+2\right\} ,\\
\sigma_{7} & =\left\{ a\neq\mu+1,a\neq\mu+2;b\neq\mu+1,b\neq\mu+2\right\} ,\\
\sigma_{8} & =\left\{ a\neq b,a\neq b'',a=\mu+1\text{ or }a=\mu+2;b\neq\mu+1,b\neq\mu+2\right\} .
\end{align}

In the equation (\ref{S}), the amplitudes $a_{k}(x)$, $1\leq k\leq8$,
are given by
\begin{align}
a_{1}(x) & =\tfrac{1}{2}q\left(x^{2}-1\right)\left(x^{2}-\zeta^{2}\right)\left(1+\kappa_{1}\right),\\
a_{2}(x) & =\tfrac{1}{2}q\left(x^{2}-1\right)\left(x^{2}-\zeta^{2}\right)\left(1-\kappa_{1}\right),\\
a_{3}(x) & =-\left(q^{2}-1\right)\left(x^{2}-\zeta^{2}\right),\\
a_{4}(x) & =-x^{2}\left(q^{2}-1\right)\left(x^{2}-\zeta^{2}\right),\\
a_{5}(x) & =\tfrac{1}{2}\left[q\left(x^{2}-1\right)\left(x^{2}-\zeta^{2}\right)\left(1+\nu\kappa_{1}\right)+x\left(x-1\right)\left(q^{2}-1\right)\left(\zeta+\kappa_{2}\right)\left(x\kappa_{2}+\zeta\right)\right],\\
a_{6}(x) & =\tfrac{1}{2}\left[q\left(x^{2}-1\right)\left(x^{2}-\zeta^{2}\right)\left(1+\nu\kappa_{1}\right)-x\left(x+1\right)\left(q^{2}-1\right)\left(\zeta+\kappa_{2}\right)\left(x\kappa_{2}-\zeta\right)\right],\\
a_{7}(x) & =\tfrac{1}{2}\left[q\left(x^{2}-1\right)\left(x^{2}-\zeta^{2}\right)\left(1-\nu\kappa_{1}\right)+x\left(x+1\right)\left(q^{2}-1\right)\left(\zeta-\kappa_{2}\right)\left(x\kappa_{2}+\zeta\right)\right],\\
a_{8}(x) & =\tfrac{1}{2}\left[q\left(x^{2}-1\right)\left(x^{2}-\zeta^{2}\right)\left(1-\nu\kappa_{1}\right)-x\left(x-1\right)\left(q^{2}-1\right)\left(\zeta-\kappa_{2}\right)\left(x\kappa_{2}-\zeta\right)\right].
\end{align}
where $\zeta=q^{n-m}$. The amplitudes $b_{k}^{a}(x)$, $1\leq k\leq5$,
which depend on the index $a$, are given by, 
\begin{align}
b_{1}^{a}(x) & =\left(x^{2}-\zeta^{2}\right)\left[x^{2\left(1-p_{a}\right)}-q^{2}x^{2p_{a}}\right],\quad1\leq a\leq N,\\
b_{2}^{a}(x) & =\begin{cases}
-\frac{1}{2}(q^{2}-1)(x^{2}-\zeta^{2})(x+1), & a<\mu+1,\\
-\frac{1}{2}x(q^{2}-1)(x^{2}-\zeta^{2})(x+1), & a>\mu+2,
\end{cases}\\
b_{3}^{a}(x) & =\begin{cases}
\frac{1}{2}(q^{2}-1)(x^{2}-\zeta^{2})(x-1), & a<\mu+1,\\
-\frac{1}{2}x(q^{2}-1)(x^{2}-\zeta^{2})(x-1), & a>\mu+2,
\end{cases}\\
b_{4}^{a}(x) & =\begin{cases}
\frac{1}{2}\left(\theta_{a}q^{\tau_{a}}\right)\left(x^{2}-1\right)\left(q^{2}-1\right)(x\kappa_{2}+\zeta), & a<\mu+1,\\
\frac{1}{2}x\left(\theta_{a}q^{\tau_{a}}\right)\left(x^{2}-1\right)\left(q^{2}-1\right)(x\kappa_{2}+\zeta), & a>\mu+2,
\end{cases}\\
b_{5}^{a}(x) & =\begin{cases}
-\frac{1}{2}\left(\theta_{a}q^{\tau_{a}}\right)\left(x^{2}-1\right)\left(q^{2}-1\right)(x\kappa_{2}-\zeta), & a<\mu+1,\\
\frac{1}{2}x\left(\theta_{a}q^{\tau_{a}}\right)\left(x^{2}-1\right)\left(q^{2}-1\right)(x\kappa_{2}-\zeta), & a>\mu+2,
\end{cases}
\end{align}
and the amplitudes $c_{k}^{ab}(x)$, $1\leq k\leq2$, that depend
on the indexes $a$ and $b$ are, 
\begin{align}
c_{1}^{ab}(x) & =\begin{cases}
(q^{2}-1)\left[\zeta^{2}(x^{2}-1)\Theta_{a,b}-\delta_{a,b''}(x^{2}-\zeta^{2})\right], & a<b,\\
(x^{2}-1)\left[(x^{2}-\zeta^{2})\left(-1\right)^{p_{a}}q^{2p_{a}}+x^{2}(q^{2}-1)\right], & a=b,\\
x^{2}(q^{2}-1)\left[(x^{2}-1)\Theta_{a,b}-\delta_{a,b''}(x^{2}-\zeta^{2})\right], & a>b,
\end{cases}\\
c_{2}^{ab}(x) & =\left(-1\right)^{p_{a}p_{b}}q\left(x^{2}-1\right)\left(x^{2}-\zeta^{2}\right),\qquad1\leq a,b\leq N.
\end{align}
We made use of the following graduation and Grassmann parity, 
\begin{equation}
p_{a}=\begin{cases}
\pi_{a}, & a<\mu+1,\\
0, & \mu+1\leq a\leq\mu+2,\\
\pi_{a-1}, & a>\mu+2,
\end{cases}\qquad\pi_{a}=\begin{cases}
1, & m+1\leq a\leq2n+m+1,\\
0, & \text{otherwise}.
\end{cases}
\end{equation}

The remaining parameters of the solution are given by,
\begin{equation}
t_{a}=\begin{cases}
-p_{a}+a+1+2{\displaystyle \sum_{b=a}^{\mu}}p_{a}, & a<\mu+1,\\
\mu+\frac{3}{2}, & \mu+1\leq a\leq\mu+2,\\
p_{a}+a-1-2{\displaystyle \sum_{b=\mu+3}^{a}}p_{b}, & a>\mu+2,
\end{cases}
\end{equation}
\begin{equation}
\tau_{a}=\begin{cases}
p_{a}+a-\frac{1}{2}-2{\displaystyle \sum_{b=a}^{\mu}}p_{b}, & a<\mu+1,\\
0, & \mu+1\leq a\leq\mu+2,\\
p_{a}+a-\frac{5}{2}-\mu-2{\displaystyle \sum_{b=\mu+3}^{a}}p_{b}, & a>\mu+2,
\end{cases}
\end{equation}
\begin{equation}
\theta_{a}=\begin{cases}
\left(-1\right)^{-p_{a}/2}, & a<\mu+1,\\
1, & \mu+1\leq a\leq\mu+2,\\
\left(-1\right)^{p_{a}/2}, & a>\mu+2,
\end{cases}\qquad\mathrm{and}\qquad\Theta_{a,b}=\frac{\theta_{a}q^{t_{a}}}{\theta_{b}q^{t_{b}}}.
\end{equation}

The $R$-matrix associated with (\ref{S}) can be obtained through
$R(x)=P_{g}S(x)$. This $R$-matrix satisfies the regularity, unitarity,
PT and crossing symmetries, which are important for the implementation
of the boundary algebraic Bethe Ansatz $-$ see \cite{Galleas2006}
for the details. 

Notice that the Galleas-Martins $S$-matrix (\ref{S}) depends on
three parameters, namely, $\kappa_{1}$, $\kappa_{2}$ and $\nu$
(we say that this $S$-matrix is \emph{multiparametric}). These parameters
can assume only the values $1$ and $-1$ and for each possibility
we get a corresponding supersymmetric vertex-model. The case $\kappa_{1}=\kappa_{2}=\nu=1$
is most important one, since it is only in this case that the $S$-matrix
(\ref{S}) reduces to Jimbo's $S$-matrix \cite{Jimbo1986A} when
the fermionic degrees of freedom are despised, \emph{i.e.}, when we
make $m=0$. Other values of $\kappa_{1}$, $\kappa_{2}$ and $\nu$
lead to other vertex-models corresponding to non-trivial generalizations
of Jimbo's $S$-matrix \cite{Jimbo1986A}.

Galleas and Martins conjectured that the symmetry behind their vertex-model
is described by the $U_{q}[\mathrm{osp}\left(2n+2|2m\right)^{\left(2\right)}]$
quantum affine Lie superalgebra \cite{Galleas2006} (see also \cite{Galleas2007}).
Their claim can be justified in the following way: first, remember
that a Lie superalgebra is defined on a $\mathbb{Z}_{2}$-graded vector
space $V$ that decomposes into the direct sum $V=V_{0}\otimes V_{1}$,
where $V_{0}$ is the even (bosonic) part of $V$ and $V_{1}$ is
its odd (fermionic) part \cite{Kac1977A,Kac1977B}. Now, since the
Galleas-Martins $S$-matrix reduces to the Jimbo's $S$-matrix \cite{Jimbo1986A}
when the fermionic degrees of freedom are despised, and since the
Jimbo $S$-matrix has the $U_{q}[\mathrm{o}^{\left(2\right)}\left(2n+2\right)]=U_{q}[D_{n+1}^{\left(2\right)}]$
symmetry, this means that the even part of the Lie superalgebra associated
with the Galleas-Martins vertex-model must have this same symmetry.
However, only the $\mathrm{osp}^{\left(2\right)}\left(2n+2|2m\right)=D^{\left(2\right)}\left(n+1|m\right)$
Lie superalgebra has an even part corresponding to the $\mathrm{o}\left(2n+2\right)^{\left(2\right)}=D_{n+1}^{\left(2\right)}$
affine Lie algebra \cite{FeingoldFrenkel1985,Frappat1989,Frappat2000,NeebPianzola2010,Serganova2011,Musson2012,Ransingh2013,Sthanumoorthy2016,Xu2016}.
In fact, we have the decomposition $\mathrm{osp}^{\left(2\right)}\left(2n+2|2m\right)=\mathrm{o}^{\left(2\right)}\left(2n+2\right)\otimes\mathrm{sp}^{\left(1\right)}\left(2m\right)$,
which in Cartan's notation becomes $D^{\left(2\right)}\left(n+1|m\right)=D_{n+1}^{\left(2\right)}\otimes C_{m}^{\left(1\right)}$
\cite{FeingoldFrenkel1985,Frappat1989,Frappat2000,NeebPianzola2010,Serganova2011,Musson2012,Ransingh2013,Sthanumoorthy2016,Xu2016}.
These decompositions are not expected to be changed as we perform
the quantum deformation of the universal enveloping algebra associated
with the $\mathrm{osp}^{\left(2\right)}\left(2n+2|2m\right)=D^{\left(2\right)}\left(n+1|m\right)$
Lie superalgebra and, hence, it follows that the Galleas-Martins vertex-model
should be associated with the $U_{q}[\mathrm{osp}^{\left(2\right)}\left(2n+2|2m\right)]=U_{q}[D^{\left(2\right)}\left(n+1|m\right)]$
quantum twisted orthosymplectic Lie superalgebra. 

For the case $n=0$ and $\kappa_{1}=\kappa_{2}=\nu=1$ we obtain a
supersymmetric vertex-model which can be thought as the fermionic
analogue of $D_{n+1}^{\left(2\right)}$ Jimbo's vertex-model \cite{Jimbo1986A}.
The underlining symmetry behind this vertex-model is the $U_{q}[\mathrm{osp}^{\left(2\right)}(2|2m)]=U_{q}[C^{\left(2\right)}\left(m+1\right)]$
quantum twisted Lie superalgebra \cite{FeingoldFrenkel1985,Frappat1989,Frappat2000,NeebPianzola2010,Serganova2011,Musson2012,Ransingh2013,Sthanumoorthy2016,Xu2016}.
We can write the $S$-matrix of the $U_{q}[\mathrm{osp}^{\left(2\right)}(2|2m)]$
vertex-model in the same form as given at (\ref{S}), with the only
changes occurring in the amplitudes, which become considerably simpler:
\begin{align}
a_{1}(x) & =q\left(x^{2}-1\right)\left(x^{2}-q^{-2m}\right),\\
a_{2}(x) & =0,\\
a_{3}(x) & =-\left(q^{2}-1\right)\left(x^{2}-q^{-2m}\right),\\
a_{4}(x) & =-x^{2}\left(q^{2}-1\right)\left(x^{2}-q^{-2m}\right),\\
a_{5}(x) & =\tfrac{1}{2}\left[2q\left(x^{2}-1\right)\left(x^{2}-q^{-2m}\right)+x\left(x-1\right)\left(q^{2}-1\right)\left(q^{-m}+1\right)\left(x+q^{-m}\right)\right],\\
a_{6}(x) & =\tfrac{1}{2}\left[2q\left(x^{2}-1\right)\left(x^{2}-q^{-2m}\right)-x\left(x+1\right)\left(q^{2}-1\right)\left(q^{-m}+1\right)\left(x-q^{-m}\right)\right],\\
a_{7}(x) & =\tfrac{1}{2}x\left(x+1\right)\left(q^{2}-1\right)\left(q^{-m}-1\right)\left(x+q^{-m}\right),\\
a_{8}(x) & =\tfrac{1}{2}\left[x\left(x-1\right)\left(q^{2}-1\right)\left(q^{-m}-1\right)\left(x-q^{-m}\right)\right],
\end{align}
\begin{align}
b_{1}^{a}(x) & =\left(x^{2}-q^{-2m}\right)\left[x^{2\left(1-p_{a}\right)}-q^{2}x^{2p_{a}}\right],\quad1\leq a\leq N,\\
b_{2}^{a}(x) & =\begin{cases}
-\frac{1}{2}\left(q^{2}-1\right)\left(x^{2}-q^{-2m}\right)\left(x+1\right), & a<m+1,\\
-\frac{1}{2}x\left(q^{2}-1\right)\left(x^{2}-q^{-2m}\right)\left(x+1\right), & a>m+2,
\end{cases}\\
b_{3}^{a}(x) & =\begin{cases}
\frac{1}{2}\left(q^{2}-1\right)\left(x^{2}-q^{-2m}\right)\left(x-1\right), & a<m+1,\\
-\frac{1}{2}x\left(q^{2}-1\right)\left(x^{2}-q^{-2m}\right)\left(x-1\right), & a>m+2,
\end{cases}\\
b_{4}^{a}(x) & =\begin{cases}
\frac{1}{2}\left(\theta_{a}q^{\tau_{a}}\right)\left(x^{2}-1\right)\left(q^{2}-1\right)\left(x+q^{-m}\right), & a<m+1,\\
\frac{1}{2}x\left(\theta_{a}q^{\tau_{a}}\right)\left(x^{2}-1\right)\left(q^{2}-1\right)\left(x+q^{-m}\right), & a>m+2,
\end{cases}\\
b_{5}^{a}(x) & =\begin{cases}
-\frac{1}{2}\left(\theta_{a}q^{\tau_{a}}\right)\left(x^{2}-1\right)\left(q^{2}-1\right)\left(x-q^{-m}\right), & a<m+1,\\
\frac{1}{2}x\left(\theta_{a}q^{\tau_{a}}\right)\left(x^{2}-1\right)\left(q^{2}-1\right)\left(x-q^{-m}\right), & a>m+2,
\end{cases}
\end{align}
and,
\begin{align}
c_{1}^{ab}(x) & =\begin{cases}
(q^{2}-1)\left[q^{-2m}(x^{2}-1)\Theta_{a,b}-\delta_{ab''}\left(x^{2}-q^{-2m}\right)\right], & a<b,\\
(x^{2}-1)\left[\left(x^{2}-q^{-2m}\right)\left(-1\right)^{p_{a}}q^{2p_{a}}+x^{2}\left(q^{2}-1\right)\right], & a=b,\\
x^{2}(q^{2}-1)\left[(x^{2}-1)\Theta_{a,b}-\delta_{ab''}\left(x^{2}-q^{-2m}\right)\right], & a>b,
\end{cases}\\
c_{2}^{ab}(x) & =\left(-1\right)^{p_{a}p_{b}}q\left(x^{2}-1\right)\left(x^{2}-q^{-2m}\right),\qquad1\leq a,b\leq N.
\end{align}

In this case, the other functions also become simpler: 
\begin{equation}
p_{a}=\begin{cases}
0, & m+1\leq a\leq m+2,\\
1, & \text{otherwise},
\end{cases}\qquad\pi_{a}=\begin{cases}
0, & a=m+1,\\
1, & \text{otherwise},
\end{cases}
\end{equation}
\begin{equation}
t_{a}=\begin{cases}
N-a, & a<m+1,\\
m+\frac{3}{2}, & m+1\leq a\leq m+2,\\
N+2-a, & a>m+2,
\end{cases}\quad\tau_{a}=\begin{cases}
\frac{1}{2}-a, & a<m+1,\\
0, & m+1\leq a\leq m+2,\\
m+\frac{5}{2}-a, & a>m+2,
\end{cases}
\end{equation}
\begin{equation}
\theta_{a}=\begin{cases}
-i, & a<m+1,\\
1, & m+1\leq a\leq m+2,\\
i, & a>m+2,
\end{cases}\qquad\mathrm{and}\qquad\Theta_{a,b}=\frac{\theta_{a}q^{t_{a}}}{\theta_{b}q^{t_{b}}}.
\end{equation}
where $i=\sqrt{-1}$. 

\section{Solutions of the boundary YB equation}

Hereafter we shall present the reflection $K$-matrices, solutions
of the boundary Yang-Baxter equations (\ref{BYB1}) or associated
to the $U_{q}[\mathrm{osp}^{\left(2\right)}\left(2|2m\right)]=U_{q}[C^{\left(2\right)}\left(m+1\right)]$
vertex-model. We shall also present the diagonal reflection $K$-matrices
associated with the $U_{q}[\mathrm{osp}^{\left(2\right)}\left(2n+2|2m\right)]=U_{q}[D^{\left(2\right)}\left(n+1|m\right)]$
vertex-model.

As commented in the previous section, the $U_{q}[\mathrm{osp^{\left(2\right)}}\left(2|2m\right)]$
vertex-model can be seen as the fermionic analogue of Jimbo's $U_{q}[\mathrm{o}^{\left(2\right)}\left(2n+2\right)]$
vertex-model \cite{Jimbo1986A}. We remark that the first solutions
of the boundary YB equation associated to the $U_{q}[\mathrm{o}^{\left(2\right)}\left(2n+2\right)]$
vertex-model were the diagonal and block-diagonal solutions found
by Martins and Guan \cite{Guan2000}; soon after Lima-Santos deduced
the general $K$-matrices of this vertex-model \cite{Lima2001}. The
corresponding reflection $K$-matrices for the multiparametric $U_{q}[\mathrm{o}^{\left(2\right)}\left(2n+2\right)]$
vertex-model were deduced and classified by Vieira and Lima-Santos
in \cite{Vieira2013}; new family of solutions for Jimbo's $U_{q}[\mathrm{o}^{\left(2\right)}\left(2n+2\right)]$
vertex-model was also derived in \cite{Vieira2013}. 

Among the graded vertex-models known up to date, the $U_{q}[\mathrm{osp}^{\left(2\right)}\left(2n+2|2m\right)]=U_{q}[D^{\left(2\right)}\left(n+1|m\right)]$
vertex-model is by far the most complex one. This can be seen either
directly from the very complexity of the $S$-matrix given at (\ref{S})
or from the highly non-trivial nature of the twisted orthosymplectic
Lie superalgebras \cite{FeingoldFrenkel1985,Frappat1989,Frappat2000,NeebPianzola2010,Serganova2011,Ransingh2013,Sthanumoorthy2016,Xu2016}.
In fact, while the reflection $K$-matrices of the $U_{q}[\mathrm{sl}^{\left(2\right)}\left(r|2m\right)]$,
$U_{q}[\mathrm{osp}^{\left(1\right)}\left(r|2m\right)]$, $U_{q}[\mathrm{spo}\left(2n|2m\right)]$
and $U_{q}[\mathrm{sl}^{\left(1\right)}\left(m|n\right)]$ vertex-models
were obtained in a period of almost one year \cite{Lima2009A,Lima2009B,Lima2009C,Lima2010},
the corresponding reflection $K$-matrices of the $U_{q}[\mathrm{osp}^{\left(2\right)}\left(2|2m\right)]$
vertex-model were derived only now, approximately eight years after.
Indeed, the general $K$-matrices for the $U_{q}[\mathrm{osp}^{\left(2\right)}\left(2n+2|2m\right)]$
vertex-model (except for the diagonal ones which we report in this
work) are yet unknown. 

The methodology used by us to solve the boundary YB equation (\ref{BYB1})
was the standard derivative method. This method was first used to
solve the periodic YB equation by Zamolodchikov and Fateev in \cite{ZamolodchikovFateev1980}
and it has been extensively used by Lima-Santos in order to solve
the boundary YB equations \cite{Lima1999,Lima2001,Malara2006,Lima2009A,Lima2009B,Lima2009C,Lima2010}. 

The derivative method consists in taking the formal derivative of
the boundary YB equation (\ref{BYB1}) with respect to one of the
variables and evaluating it at some particular value of that chosen
variable. For instance, taking the formal derivative of (\ref{BYB1})
with respect to $y$ and evaluating the resulting expression at $y=1$
we shall get the equation
\begin{multline}
2R_{12}\left(x\right)K_{1}\left(x\right)D_{21}\left(x\right)+2D_{12}\left(x\right)K_{1}\left(x\right)R_{21}\left(x\right)+\\
R_{12}\left(x\right)K_{1}\left(x\right)R_{21}\left(x\right)B_{2}-B_{2}R_{12}\left(x\right)K_{1}\left(x\right)R_{21}\left(x\right)=0,\label{DBYBE}
\end{multline}
where, 
\begin{equation}
D_{12}\left(x\right)=\left.\frac{\partial R_{12}\left(x/y\right)}{\partial y}\right|_{y=1},\quad D_{21}\left(x\right)=\left.\frac{\partial R_{21}\left(x/y\right)}{\partial y}\right|_{y=1},\quad B_{2}=\left.\frac{dK_{2}(y)}{dy}\right|_{y=1}.
\end{equation}
(we have used the fact that the reflection $K$-matrix is regular,
which means that it satisfies the property $K\left(1\right)=I$).
This procedure\footnote{In the case of the periodic YB equation, the derivative method leads
to a system of differential equations instead of algebraic equations.
This is due to the fact that the $R$-matrices appearing on the periodic
YB equation depends on two variables instead of one variable.} allows us to convert the set of $N^{4}$ non-linear functional equations
(\ref{BYB1}), which depends on the two unknowns $x$ and $y$, into
a set of $N^{4}$ linear functional equations depending only on the
variable $x$. This, however, comes with a price: the introduction
of a set of $N^{2}$ \emph{boundary parameters} 
\begin{equation}
\beta_{a.b}=\left.\frac{dk_{a,b}(y)}{dy}\right|_{y=1},\qquad1\leq a,b\leq N.\label{Derivative}
\end{equation}

We remark that although the system (\ref{DBYBE}) is overdetermined,
it is nevertheless consistent. This remarkable property is due to
the existence of the additional boundary parameters $\beta_{a,b}$,
$1\leq a,b\leq N$, which allow to solve the remaining functional
equations, after all elements of the $K$-matrix are determined as
functions of these boundary parameters. Actually, in general we need
to fix only a subset of the boundary parameters $\beta_{a,b}$ in
order to solve all the functional equations of the system (\ref{DBYBE});
the remaining boundary parameters that did not need to be fixed are
the \emph{boundary free-parameters} of the solution). 

Although the \emph{derivative boundary YB equation} (\ref{DBYBE})
consists in a linear system of functional equations depending only
on the variable $x$, this system is still very difficult to be solved
$-$ in fact, even just writing the $R$-matrix and verifying that
it satisfies the YB equations by itself tough task. Moreover, the
complexity of the system is very sensitive to the order on which the
equations are solved and on what elements of the $K$-matrix are eliminated
first. An unfortunate choice for solving the equations generally increases
the complexity of the system in such a way that even with the most
powerful computational resources the solution could not be achieved.

In the following, we shall describe a recipe for a possible order
of solving the system of functional equations (\ref{DBYBE}) on which
the complexity of the system can be maintained under control and whence
their solutions can be found.
\begin{enumerate}
\item The simplest equations of the derivative boundary YB equation (\ref{DBYBE})
are those containing only the non-diagonal elements of the reflection
$K$-matrix (different from $k_{m+1,m+2}\left(x\right)$ and $k_{m+2,m+1}\left(x\right)$)
not lying on its first or last line or column. We can use these equations
to eliminate the elements $k_{a,b}\left(x\right)$, $1<a,b<N$, $\left\{ a,b\right\} \neq\left\{ m+1,m+2\right\} $,
in favor of the elements $k_{1,b}\left(x\right)$, $1<b<N$, and $k_{a,1}\left(x\right)$,
$1<a<N$.
\item Next we should look for those equations containing only the elements
lying on the first or last line or column. In this way we can eliminate
the elements $k_{1,b}\left(x\right)$, $1<b<N$, in terms of $k_{1,N}\left(x\right)$
and the elements $k_{a,1}\left(x\right)$, $1<a<N$, in terms of $k_{N,1}\left(x\right)$.
\item Now we can search for equation containing only elements lying on the
secondary diagonal of the reflection $K$-matrix. We can solve these
equations in favor of the elements $k_{a,N+1-a}\left(x\right)$, $2\leq a\leq m$,
in terms of $k_{1,N}\left(x\right)$ and the elements $k_{a,N+1-a}\left(x\right)$,
$m+3\leq a\leq N$, in terms of $K_{N,1}\left(x\right)$. 
\item Other equations containing only $K_{N,1}\left(x\right)$ and $K_{1,N}\left(x\right)$
will provide the expression of $K_{N,1}\left(x\right)$ in terms of
$K_{1,N}\left(x\right)$.
\item At this point, the system becomes very complex and the remaining expressions
for the reflection $K$-matrices elements pass to depend on the parity
of $N$. Notwithstanding the high complexity of the system, we can
find equations that provide the diagonal elements $k_{a,a}\left(x\right)$,
$2\leq a\leq m$, in terms of $k_{1,1}\left(x\right)$ and $k_{1,N}\left(x\right)$
and the diagonal elements containing $k_{a,a}\left(x\right)$, $m+3\leq a\leq N$
in terms of $k_{m+3,m+3}\left(x\right)$ and $k_{1,N}\left(x\right)$. 
\item Then we can find the expressions of $k_{m+3,m+3}\left(x\right)$ and
$k_{1,1}\left(x\right)$ in terms of $k_{1,N}\left(x\right)$. These
diagonal elements will satisfy welcome recurrence relations.
\item Provided the computer machine has sufficient power to handle the equations,
the remaining central elements $k_{m+1,m+1}\left(x\right)$, $k_{m+1,m+2}\left(x\right)$,
$k_{m+2,m+1}\left(x\right)$ and $k_{m+2,m+2}\left(x\right)$ can
be eliminated in terms of $k_{1,N}\left(x\right)$.
\item At this point all elements of the reflection $K$-matrix will be eliminated
in terms of the element $k_{1,N}\left(x\right)$. Then, we can give
to $k_{1,N}\left(x\right)$ any desirable value so that if satisfies
the properties $k_{1,N}\left(1\right)=0$ and $k_{1,N}'\left(0\right)=\beta_{1,N}$. 
\item Although all elements of the reflection $K$-matrix are determined
as functions of $x$, $q$ and the boundary parameters $\beta_{a,b}$,
we can verify that several functional equations still are not satisfied.
In order to solve these remaining equations, a sufficient number of
constraints between the boundary parameters $\beta_{a,b}$ should
be found. As doing so, the solution may present branches if some quadratic
(or of high degree) expressions for the boundary parameters appears.
Every branch must be carefully taken into account in order to no solution
be missed.
\item Finally we must check if the solution is regular and it derivative
is in accordance with the definition of the boundary parameters given
at (\ref{Derivative}). If these properties are not yet satisfied,
further boundary parameters should be fixed until the solution becomes
regular and consistent. after this we are done and we shall have the
solution of the problem. 
\end{enumerate}
Once we have the solution of the derivative boundary YB equation (\ref{DBYBE}),
we can verify that the reflection $K$-matrix are indeed solutions
of the boundary YB equation (\ref{BYB1}). We would like to emphasize
that the intermediary expressions for the reflection $K$-matrix elements
(and the reflection equation as well) that appear as we solve the
equations are extremely huge and, as a matter of a fact, not important
at all. By this reason we shall write in the sequel only the final
expressions for the reflection $K$-matrix elements.

We classified the reflection $K$-matrices for the $U_{q}[\mathrm{osp}^{\left(2\right)}\left(2|2m\right)]=U_{q}[C^{\left(2\right)}\left(m+1\right)]$
vertex-model into four classes, as described below:
\begin{itemize}
\item \textbf{Complete solutions:} These are the most general solutions
we found, where no element of the $K$-matrix is null. These solutions
are characterized by $m$ boundary free-parameters for a given $N$
and we found one family of solutions that branches into two subfamilies
differing by the value of $\epsilon=\pm1$.
\item \textbf{Block-diagonal solutions:} These are solutions on which the
reflection $K$-matrices are almost diagonal: all non-diagonal elements,
excepting the elements $k_{m+1,m+2}(x)$ and $k_{m+2,m+1}(x)$, are
null. The shape of this matrix is related to the existence of $m$
distinct conserved $U(1)$ charges \cite{Guan2000,Lima2001}. We found
two families of block-diagonal solutions, which are characterized
by only one boundary free-parameter. Each family also branches into
two subfamilies differing by the value of $\epsilon=\pm1$. 
\item \textbf{\emph{X}}\textbf{-shape solutions:} In this case the only
non-vanishing elements of the reflection $K$-matrices are those lying
on the main and the secondary diagonals. We found only one family
of \emph{X}-shape solutions that, for a given $N$, contain $m$ boundary
free-parameters. There is no branch here.
\item \textbf{Diagonal Solutions:} Finally, we found two families of diagonal
solutions which are actually valid for the $U_{q}[\mathrm{osp}^{\left(2\right)}\left(2n+2|2m\right)]=U_{q}[D^{\left(2\right)}\left(n+1|m\right)]$
vertex-model. The first family of diagonal reflection $K$-matrices
holds for any values of $m$ and $n$ and has no free-parameter. The
second family holds only when $m=n$ and has two free-parameters.
\end{itemize}
Besides the solutions commented above, we present in the appendix
two particular families of solutions which hold only for the $U_{q}[\mathrm{osp}^{\left(2\right)}\left(2|2\right)]=U_{q}[C^{\left(2\right)}\left(2\right)]$
and $U_{q}[\mathrm{osp}^{\left(2\right)}\left(2|4\right)]=U_{q}[C^{\left(2\right)}\left(3\right)]$
vertex-models, respectively. 

\subsection{Complete solutions \label{subsec:The-main-solution}}

The complete solutions are the most general reflection $K$-matrices
we found. In this case, all elements of the $K$-matrix are different
from zero. The solutions present two branches determined by $\epsilon=\pm1$
and they are characterized by $m$ free parameters, namely, $\beta_{1,m+2},\beta_{1,m+3}\ldots,\beta_{1,N-2}$
and $\beta_{1,N-1}$. 

We begin by defining the quantities 
\begin{equation}
\beta_{\pm}=\frac{1}{2}\left(\beta_{1,m+1}\pm\beta_{1,m+2}\right),\qquad G_{m}(x)=\frac{q^{1-m}+1}{q^{1-m}+x^{2}},\qquad H_{m}=\frac{q^{1-m}+1}{q+1}.
\end{equation}

With the help of this quantities, we can write the elements of the
$K$-matrix as follows: . for the first line of the reflection $K$-matrix,
we have, 
\begin{equation}
k_{1,m+1}(x)=\left(\frac{\beta_{+}+x\beta_{-}}{\beta_{1,N}}\right)G_{m}(x)k_{1,N}(x),
\end{equation}
\begin{equation}
k_{1,m+2}(x)=\left(\frac{\beta_{+}-x\beta_{-}}{\beta_{1,N}}\right)G_{m}(x)k_{1,N}(x),
\end{equation}
\begin{equation}
k_{1,j}(x)=\left(\frac{\beta_{1,b}}{\beta_{1,N}}\right)G_{m}(x)k_{1,N}(x),\qquad1<b<N,b\neq m+1,b\neq m+2,
\end{equation}
and, its first column, we have 
\begin{equation}
k_{m+1,1}(x)=\Theta_{m+1,2}\left(\frac{\beta_{2,1}}{\beta_{1,N-1}}\right)\left(\frac{\beta_{+}-x\beta_{-}}{\beta_{1,N}}\right)G_{m}(x)k_{1,N}(x),
\end{equation}
\begin{equation}
k_{m+2,1}(x)=\Theta_{m+2,2}\left(\frac{\beta_{2,1}}{\beta_{1,N-1}}\right)\left(\frac{\beta_{+}+x\beta_{-}}{\beta_{1,N}}\right)G_{m}(x)k_{1,N}(x),
\end{equation}
\begin{multline}
k_{a,1}(x)=\Theta_{a,2}\left(\frac{\beta_{2,1}}{\beta_{1,N-1}}\right)\left(\frac{\beta_{1,a'}}{\beta_{1,N}}\right)G_{m}(x)k_{1,N}(x),\\
1<a<N,a\neq m+1,a\neq m+2.
\end{multline}
For the elements of the last line, we have, 
\begin{equation}
k_{N,m+1}(x)=xq^{m}\Theta_{N,2}\left(\frac{\beta_{2,1}}{\beta_{1,N-1}}\right)\left(\frac{x\beta_{+}+q^{-m}\beta_{-}}{\beta_{1,N}}\right)G_{m}(x)k_{1,N}(x),
\end{equation}
\begin{equation}
k_{N,m+2}(x)=xq^{m}\Theta_{N,2}\left(\frac{\beta_{2,1}}{\beta_{1,N-1}}\right)\left(\frac{x\beta_{+}-q^{-m}\beta_{-}}{\beta_{1,N}}\right)G_{m}(x)k_{1,N}(x),
\end{equation}
\begin{multline}
k_{N,b}(x)=x^{2}q^{m}\Theta_{N,2}\left(\frac{\beta_{2,1}}{\beta_{1,N-1}}\right)\left(\frac{\beta_{1,b}}{\beta_{1,N}}\right)G_{m}(x)k_{1,N}(x),\\
1<b<N,b\neq m+1,b\neq m+2,
\end{multline}
and, for those in the last column, 
\begin{equation}
k_{m+1,N}(x)=xq^{m}\Theta_{m+1,1}\left(\frac{x\beta_{+}-q^{-m}\beta_{-}}{\beta_{1,N}}\right)G_{m}(x)k_{1,N}(x),
\end{equation}
\begin{equation}
k_{m+2,N}(x)=xq^{m}\Theta_{m+2,1}\left(\frac{x\beta_{+}+q^{-m}\beta_{-}}{\beta_{1,N}}\right)G_{m}(x)k_{1,N}(x),
\end{equation}
\begin{equation}
k_{a,N}(x)=x^{2}q^{m}\Theta_{a,1}\left(\frac{\beta_{1,a'}}{\beta_{1,N}}\right)G_{m}(x)k_{1,N}(x),\qquad1<a<N,a\neq m+1,a\neq m+2.
\end{equation}

For the elements lying on the secondary diagonal not in the center
of $K$-matrix (\emph{i.e.,} for $a\neq m+1$ and $a\neq m+2$) we
have, 
\begin{align}
k_{1,N}(x) & =\frac{1}{2}\left(x^{2}-1\right)\beta_{1,N},
\end{align}
\begin{equation}
k_{N,1}(x)=\Theta_{N-1,2}\left(\frac{\beta_{2,1}}{\beta_{1,N-1}}\right)^{2}k_{1,N}(x),
\end{equation}
\begin{multline}
k_{a,a'}\left(x\right)=\left(-1\right)^{p_{a}}q\Theta_{1,a'}\left(\frac{\beta_{1,a'}}{\beta_{1,N}}\right)^{2}H_{m}^{2}k_{1,N}(x),\\
1<a<N,a\neq m+1,a\neq m+2,
\end{multline}
and for the elements of the $K$-matrix above the secondary diagonal,
not in the first line or in the first column, we have
\begin{multline}
k_{m+1,b}(x)=q^{m}\Theta_{m+1,1}\left(\frac{\beta_{1,b}}{\beta_{1,N}}\right)\left(\frac{\beta_{+}-x\beta_{-}}{\beta_{1,N}}\right)H_{m}G_{m}(x)k_{1,N}(x),\\
b\neq m+1,b\neq m+2
\end{multline}
\begin{multline}
k_{m+2,b}(x)=q^{m}\Theta_{m+2,1}\left(\frac{\beta_{1,b}}{\beta_{1,N}}\right)\left(\frac{\beta_{+}+x\beta_{-}}{\beta_{1,N}}\right)H_{m}G_{m}(x)k_{1,N}(x),\\
b\neq m+1,b\neq m+2
\end{multline}
\begin{multline}
k_{a,m+1}(x)=q^{m}\Theta_{a,1}\left(\frac{\beta_{1,a'}}{\beta_{1,N}}\right)\left(\frac{\beta_{+}+x\beta_{-}}{\beta_{1,N}}\right)H_{m}G_{m}(x)k_{1,N}(x),\\
a\neq m+1,a\neq m+2,
\end{multline}
\begin{multline}
k_{a,m+2}(x)=q^{m}\Theta_{a,1}\left(\frac{\beta_{1,a'}}{\beta_{1,N}}\right)\left(\frac{\beta_{+}-x\beta_{-}}{\beta_{1,N}}\right)H_{m}G_{m}(x)k_{1,N}(x),\\
a\neq m+1,a\neq m+2,
\end{multline}
\begin{multline}
k_{a,b}(x)=q^{m}\Theta_{a,1}\left(\frac{\beta_{1,a'}}{\beta_{1,N}}\right)\left(\frac{\beta_{1,b}}{\beta_{1,N}}\right)H_{m}G_{m}(x)k_{1,N}(x),\\
a\neq m+1,a\neq m+2,j\neq m+1,j\neq m+2.
\end{multline}
Finally, for the elements below the secondary diagonal, not in the
last line or column, we have, 
\begin{multline}
k_{m+1,b}(x)=xq^{2m}\Theta_{m+1,1}\left(\frac{\beta_{1,b}}{\beta_{1,N}}\right)\left(\frac{x\beta_{+}-q^{-m}\beta_{-}}{\beta_{1,N}}\right)H_{m}G_{m}(x)k_{1,N}(x),\\
b\neq m+1,b\neq m+2
\end{multline}
\begin{multline}
k_{m+2,b}(x)=xq^{2m}\Theta_{m+2,1}\left(\frac{\beta_{1,b}}{\beta_{1,N}}\right)\left(\frac{x\beta_{+}+q^{-m}\beta_{-}}{\beta_{1,N}}\right)H_{m}G_{m}(x)k_{1,N}(x),\\
b\neq m+1,b\neq m+2
\end{multline}
\begin{multline}
k_{a,m+1}(x)=xq^{2m}\Theta_{a,1}\left(\frac{\beta_{1,a'}}{\beta_{1,N}}\right)\left(\frac{x\beta_{+}+q^{-m}\beta_{-}}{\beta_{1,N}}\right)H_{m}G_{m}(x)k_{1,N}(x),\\
a\neq m+1,a\neq m+2,
\end{multline}
\begin{multline}
k_{a,m+2}(x)=xq^{2m}\Theta_{a,1}\left(\frac{\beta_{1,a'}}{\beta_{1,N}}\right)\left(\frac{x\beta_{+}-q^{-m}\beta_{-}}{\beta_{1,N}}\right)H_{m}G_{m}(x)k_{1,N}(x),\\
a\neq m+1,a\neq m+2,
\end{multline}
\begin{multline}
k_{a,b}(x)=x^{2}q^{2m}\Theta_{a,1}\left(\frac{\beta_{1,a'}}{\beta_{1,N}}\right)\left(\frac{\beta_{1,b}}{\beta_{1,N}}\right)H_{m}G_{m}(x)k_{1,N}(x),\\
a\neq m+1,a\neq m+2,b\neq m+1,b\neq m+2.
\end{multline}

The other elements of the $K$-matrix depend on the parity of $m$
and hence, it is convenient to introduce the notation $\sigma_{m}=\left(-1\right)^{m}$.
It follows that the elements on the center of the $K$-matrix are
given by, 
\begin{equation}
k_{m+2,m+2}\left(x\right)=k_{m+1,m+1}\left(x\right)\qquad\text{and}\qquad k_{m+2,m+1}\left(x\right)=k_{m+1,m+2}\left(x\right),
\end{equation}
where, 
\begin{multline}
k_{m+1,m+1}\left(x\right)=x^{2}G_{m}(x)\left\{ \frac{\left(\sigma_{m}+1\right)}{2}\right.\\
\left.-\left(\sigma_{m}-1\right)\left[\frac{x^{2}q^{m}\left[\left(1-x^{4}\right)q+\left(q^{2}-1\right)\right]-\left(qx^{2}+1\right)\left(q-x^{2}\right)}{x^{2}\left(x^{2}+1\right)\left(q^{m}-1\right)\left(q^{2}-1\right)}\right]\right\} 
\end{multline}
and 
\begin{multline}
k_{m+1,m+2}\left(x\right)=\epsilon x^{2}G_{m}(x)\left\{ \frac{\left(\sigma_{m}-1\right)}{2}\left[\left(\frac{x^{2}-1}{x^{2}+1}\right)\left(\frac{q^{m}+1}{q^{m}-1}\right)\right]\right.\\
\left.+\frac{\left(\sigma_{m}+1\right)}{2}\left[1-\left(\frac{x^{2}q^{m}-1}{q^{m}-1}\right)\left(\frac{x^{2}q+1}{q^{2}-1}\right)\left(\frac{x^{2}+q}{x^{2}+1}\right)\right]\right\} ,
\end{multline}
with $\epsilon=\pm1$ representing two branches of the solutions.

By its turn, the diagonal elements are given recursively by 
\begin{equation}
k_{a,a}(x)=\begin{cases}
{\displaystyle k_{a-1,a-1}(x)+\left(\frac{\beta_{a,a}-\beta_{a-1,a-1}}{\beta_{1,N}}\right)G_{m}(x)k_{1,N}(x),} & 1<a<m+1\\
{\displaystyle k_{a-1,a-1}(x)+\left(\frac{\beta_{a,a}-\beta_{a-1,a-1}}{\beta_{1,N}}\right)x^{2}G_{m}(x)k_{1,N}(x),} & m+3<a<N,
\end{cases}
\end{equation}
with 
\begin{multline}
k_{1,1}\left(x\right)=G_{m}(x)\left\{ \left(\frac{x^{2}q^{m}-1}{q^{m}-1}\right)\left(\frac{q+\sigma_{m}}{q-1}\right)-\left(\frac{1+\sigma_{m}}{q-1}\right)\right.\\
-\epsilon\left(\frac{x^{2}q^{m}-1}{q^{m}-1}\right)\left(\frac{x^{2}-1}{x^{2}+1}\right)\left(\frac{q-\sigma_{m}}{q-1}\right)\\
\left.+\left(\frac{q^{m}+1}{q^{m}-1}\right)\left(\frac{\sigma_{m}-1}{q-1}\right)\left(\frac{x^{2}+\sigma_{m}}{x^{2}+1}\right)\right\} ,
\end{multline}
 and
\begin{multline}
k_{m+3,m+3}\left(x\right)=x^{2}G_{m}(x)\left\{ \left(\frac{x^{2}q^{m}-1}{q^{m}-1}\right)\left(\frac{q+\sigma_{m}}{q-1}\right)-\left(\frac{1+\sigma_{m}}{q-1}\right)\right.\\
+\epsilon\left(\frac{x^{2}q^{m}-1}{q^{m}-1}\right)\left(\frac{x^{2}-1}{x^{2}+1}\right)\left(\frac{q-\sigma_{m}}{q-1}\right)\\
\left.-\epsilon\left(\frac{q^{m}+1}{q^{m}-1}\right)\left(\frac{\sigma_{m}-1}{q-1}\right)\left(\frac{x^{2}+\sigma_{m}}{x^{2}+1}\right)\right\} .
\end{multline}

At this point all elements of the $K$-matrix were determined, but
not all functional equations are indeed satisfied. To solve the remaining
functional equations it is necessary to fix some of the parameters
$\beta_{a,b}$. The necessary and sufficient constraints between these
parameters are provided by the remaining functional equations. In
fact, these equations enable us to fix the diagonal parameters $\beta_{a,a}$
according to the recursive relations 
\begin{equation}
\beta_{a,a}=\begin{cases}
{\displaystyle \beta_{a-1,a-1}+\frac{2\sigma_{m}\left(-1\right)^{a}\left[\epsilon\left(\sigma_{m}-1\right)-\left(\sigma_{m}+1\right)\right]q^{m+1-a}\left(q+1\right)}{\left(q^{m}-1\right)\left(q-1\right)},}\\
1<a<m+1,\\
{\displaystyle \beta_{a-1,a-1}+\frac{2\sigma_{m}\left(-1\right)^{a}\left[\epsilon\left(\sigma_{m}-1\right)-\left(\sigma_{m}+1\right)\right]q^{N+1-a}\left(q+1\right)}{\left(q^{m}-1\right)\left(q-1\right)},}\\
1<a<m+1,
\end{cases}
\end{equation}
and also the following non-diagonal parameters, 
\begin{align}
\beta_{1,m+1} & =\epsilon\beta_{1,m+2},\label{beta1m}\\
\beta_{2,1} & =4iq^{2m-3/2}\left(\frac{\epsilon\left(q^{m}+1\right)+\left(q^{m}-1\right)}{\left(q-1\right)\left(q^{m}-1\right)^{2}}\right)\frac{\beta_{1,N-1}}{\beta_{1,m+2}^{2}},\label{beta21}\\
\beta_{1,N} & =-\frac{i}{4}\left\{ \frac{\left[\epsilon\left(q^{m}+1\right)+\left(q^{m}-1\right)\right]\left(q^{m}-1\right)\left(q^{m-1}+1\right)}{\left(q+1\right)q^{2m-3/2}}\right\} \beta_{1,m+2}^{2},\label{beta1N}
\end{align}
and 
\begin{multline}
\beta_{1,b}=\frac{i}{2}\left[\epsilon\left(\sigma_{m}-1\right)+\left(\sigma_{m}+1\right)\right]\left(-1\right)^{b}\left(\frac{\epsilon\left(q^{m}+1\right)+\left(q^{m}-1\right)}{q^{m-1/2}\left(q-1\right)}\right)\frac{\beta_{1,m+2}^{2}}{\beta_{1,b'}},\\
1<b<m+1.\label{beta1b}
\end{multline}

Once these parameters above are fixed, we can verify that all functional
equations are satisfied. The following $m$ parameters $\beta_{1,m+2},{}_{1,m+3}\ldots,\beta_{1,N-2}$
and $\beta_{1,N-1}$ remains arbitrary $-$ they are the free parameters
of the solution. The other parameters $\beta_{a,b}$ can be directly
found by (\ref{Derivative}) but since they do not appear explicitly
in the solution, it is not necessary write down their expressions.
The solution thus obtained is regular and characterized by $m$ free-parameters.

\subsection{Block-diagonal Solutions}

The block-diagonal solutions are such that the only non-diagonal elements
of the $K$-matrix different from zero are the elements $k_{m+1,m+2}\left(x\right)$
and $k_{m+1,m+2}\left(x\right)$. These are not reductions of the
complete solution presented in the previous section. The existence
of these block-diagonal solutions are related to the existence of
$m$ distinct conserved $U(1)$ charges and the $K$-matrix associated
to this symmetry is of the block-diagonal shape \cite{Guan2000,Lima2001}. 

We found here two families of block-diagonal solutions, each of them
branching into two solutions regarding the values of $\epsilon$.
Hence we get four families of block-diagonal solutions. These solutions
contain only one free parameter, which we choose to be $\beta_{m+1,m+2}$. 

\subsubsection{The first family of block-diagonal solutions}

For the first family of block-diagonal solutions we have that, 
\begin{equation}
k_{m+2,m+1}(x)=k_{m+1,m+2}(x)=\frac{1}{2}x^{2}(x^{2}-1)\beta_{m+1,m+2}.
\end{equation}
The other two elements lying on the center of the $K$-matrix are
given respectively by 
\begin{equation}
k_{m+1,m+1}(x)=\frac{x^{2}\left(x^{2}+1\right)}{2}+\epsilon\frac{x(x^{4}-1)}{2}\frac{q^{m/2}}{(q^{m}+1)}\sqrt{\left(\frac{q^{m}+1}{q^{m}-1}\right)^{2}\beta_{m+1,m+2}^{2}-1},
\end{equation}
and 
\begin{equation}
k_{m+2,m+2}(x)=\frac{(x^{2}+1)}{2}-\epsilon\frac{x(x^{4}-1)}{2}\frac{q^{m/2}}{(q^{m}+1)}\sqrt{\left(\frac{q^{m}+1}{q^{m}-1}\right)^{2}\beta_{m+1,m+2}^{2}-1}.
\end{equation}

Finally, the diagonal elements not in the center are given by 
\begin{multline}
k_{a,a}(x)=\frac{1}{2}\left(\frac{q^{m}x^{2}+1}{q^{2m}-1}\right)\left[(x^{2}+1)(q^{m}-1)+(x^{2}-1)(q^{m}+1)\beta_{m+1,m+2}\right],\\
1\leq a\leq m,
\end{multline}
 and 
\begin{multline}
k_{a,a}(x)=\frac{x^{2}}{2}\left(\frac{q^{m}x^{2}+1}{q^{2m}-1}\right)\left[(x^{2}+1)(q^{m}-1)-(x^{2}-1)(q^{m}+1)\beta_{m+1,m+2}\right],\\
m+3\leq a\leq N.
\end{multline}

\subsubsection{The second family of block-diagonal solutions}

For the second family of block-diagonal solutions we have, now, 
\begin{equation}
k_{m+2,m+2}(x)=k_{m+1,m+1}(x)=\frac{1}{2}x^{2}(x^{2}+1).
\end{equation}
The other two elements in the center are, 
\begin{multline}
k_{m+1,m+2}\left(x\right)=\frac{1}{2}x\left(x^{2}-1\right)\left\{ \left[\frac{x\left(q^{m}+1\right)^{2}-2q^{m}\left(x^{2}+1\right)}{\left(q^{m}-1\right)^{2}}\right]\beta_{m+1,m+2}\right.\\
\left.-\epsilon\left(x-1\right)^{2}q^{m/2}\left(\frac{q^{m}+1}{q^{m}-1}\right)^{2}\sqrt{\beta_{m+1,m+2}^{2}-1}\right\} ,
\end{multline}
 and 
\begin{multline}
k_{m+2,m+1}\left(x\right)=\frac{1}{2}x\left(x^{2}-1\right)\left\{ \left[\frac{x\left(q^{m}+1\right)^{2}+2q^{m}\left(x^{2}+1\right)}{\left(q^{m}-1\right)^{2}}\right]\beta_{m+1,m+2}\right.\\
\left.+\epsilon\left(x-1\right)^{2}q^{m/2}\left(\frac{q^{m}+1}{q^{m}-1}\right)^{2}\sqrt{\beta_{m+1,m+2}^{2}-1}\right\} .
\end{multline}
Finally, the diagonal elements are 
\begin{multline}
k_{a,a}\left(x\right)=\frac{1}{2}\frac{\left(q^{m}x^{2}-1\right)}{\left(q^{m}-1\right)^{2}}\left\{ \left(x^{2}+1\right)\left(q^{m}-1\right)+\left(x^{2}-1\right)\left(q^{m}+1\right)\beta_{m+1,m+2}\right.,\\
\left.+2\epsilon\left(x^{2}-1\right)q^{m/2}\sqrt{\beta_{m+1,m+2}^{2}-1}\right\} ,\qquad1\leq a\leq m,
\end{multline}
 and
\begin{multline}
k_{a,a}\left(x\right)=\frac{x^{2}}{2}\frac{\left(q^{m}x^{2}-1\right)}{\left(q^{m}-1\right)^{2}}\left\{ \left(x^{2}+1\right)\left(q^{m}-1\right)+\left(x^{2}-1\right)\left(q^{m}+1\right)\beta_{m+1,m+2}\right.,\\
\left.-2\epsilon\left(x^{2}-1\right)q^{m/2}\sqrt{\beta_{m+1,m+2}^{2}-1}\right\} ,\qquad m+3\leq a\leq N.
\end{multline}

\subsection{\emph{X}-shape Solutions}

There is an interesting family of solutions in which the $K$-matrix
has a shape of the letter \emph{X}. This means that the only non-null
elements of the $K$-matrix are those lying on the main or in the
secondary diagonals. Notice that in this case all bosonic degrees
of freedoms are null.

In this family of solutions, the elements lying on the main diagonal
are given by 
\begin{equation}
k_{a,a}(x)=\begin{cases}
1, & 1\leq a\leq m,\\
{\displaystyle \frac{qx^{2}+1}{q+1},} & m+1\leq a\leq m+2,\\
x^{2}, & m+3\leq a\leq N,
\end{cases}
\end{equation}
 while the elements of the secondary diagonal are, 
\begin{equation}
k_{a,a'}(x)=\begin{cases}
{\displaystyle \tfrac{1}{2}\left(x^{2}-1\right)\beta_{a,a'}}, & 1\leq a\leq m,\\
0, & m+1\leq a\leq m+2,\\
{\displaystyle \tfrac{1}{2}\left(x^{2}-1\right)\beta_{a,a'}}, & m+2\leq a\leq N.
\end{cases}
\end{equation}
The parameters $\beta_{a,a'}$ should satisfy by the constraints 
\begin{equation}
\beta_{a,a'}\beta_{a',a}=\frac{4q}{\left(q-1\right)^{2}},\qquad1\leq a\leq m,
\end{equation}
in order to all functional equations be satisfied. Whence, we get
a solution with $m$ free-parameters, namely, $\beta_{m,m+3},\beta_{2,N-1},\ldots,\beta_{1,N-1}$
and $\beta_{1,N}$.

\subsection{Diagonal Solutions for the $U_{q}[\mathrm{osp}^{\left(2\right)}\left(2n+2|2m\right)]=U_{q}[D^{\left(2\right)}\left(n+1|m\right)]$
vertex-model}

The diagonal solutions presented here are indeed valid for the $U_{q}[\mathrm{osp}^{\left(2\right)}\left(2n+2|2m\right)]=U_{q}[D^{\left(2\right)}\left(n+1|m\right)]$
vertex-model. We should remark that these diagonal solutions were
the only solutions found by us for the case $n\neq0$ so far. The
problem of finding the non-diagonal reflection $K$-matrices for the
$U_{q}[\mathrm{osp}^{\left(2\right)}\left(2n+2|2m\right)]=U_{q}[D^{\left(2\right)}\left(n+1|m\right)]$
vertex-models had eluded us so far. We intend to analyze this issue
in the future.

We found two families of diagonal solutions for the $U_{q}[\mathrm{osp}^{\left(2\right)}\left(2n+2|2m\right)]=U_{q}[D^{\left(2\right)}\left(n+1|m\right)]$
vertex-model with no free-parameters. The first one is valid for any
value of $m$ and $n$, and has no boundary free-parameter. It is
given by 
\begin{equation}
k_{a,a}(x)=\begin{cases}
1, & 1\leq a\leq m+n,\\
{\displaystyle x\left(\frac{x+q^{m+n}+i\epsilon(x-1)q^{\left(m+n\right)/2}}{1+xq^{m+n}-i\epsilon\left(x-1\right)q^{\left(m+n\right)/2}}\right),} & a=m+n+1,\\
{\displaystyle x\left(\frac{x-q^{m+n}+i\epsilon(x+1)q^{\left(m+n\right)/2}}{1-xq^{m+n}+i\epsilon\left(x+1\right)q^{\left(m+n\right)/2}}\right),} & a=m+n+2,\\
x^{2}, & m+n+3\leq a\leq N,
\end{cases}
\end{equation}
The second family of diagonal $K$-matrices holds actually only when
$n=m$. In this case, the solution has two boundary free-parameters
and it is given by, 
\begin{equation}
k_{a,a}(x)=\begin{cases}
1+\left(x-1\right)\beta_{1,1}, & 1\leq a\leq2m,\\
{\displaystyle x\varPhi_{m}^{+}\left(x\right)\left[1+\left(x-1\right)\beta_{1,1}\right],} & a=2m+1,\\
{\displaystyle x\varPhi_{m}^{-}\left(x\right)}\left[1+\left(x-1\right)\beta_{1,1}\right], & a=2m+2,\\
x^{2}\varPhi_{m}^{+}\varPhi_{m}^{-}\left[1+\left(x-1\right)\beta_{1,1}\right], & 2m+3\leq a\leq N,
\end{cases}
\end{equation}
 where, 
\begin{equation}
\varPhi_{m}^{\pm}=\frac{2\pm\left(\beta_{2m+1}-\beta_{1,1}\right)\left(x-1\right)}{2x-\left(\beta_{2m+1}-\beta_{1,1}\right)\left(x-1\right)}.
\end{equation}

\section{Conclusion}

In this work we presented the reflection $K$-matrices for the $U_{q}[\mathrm{osp^{\left(2\right)}}\left(2|2m\right)]=U_{q}[C^{\left(2\right)}\left(m+1\right)]$
vertex-model. We found several families of solutions which can be
classified into four classes: complete solutions, block-diagonal solutions,
\emph{X}-shape solutions and diagonal solutions. These diagonal solutions
are indeed valid for the $U_{q}[\mathrm{osp}^{\left(2\right)}\left(2n+2|2m\right)]=U_{q}[D^{\left(2\right)}\left(n+1|m\right)]$
vertex-model. Some special solutions which are valid only for the
$U_{q}[\mathrm{osp}^{\left(2\right)}\left(2|2\right)]$ and $U_{q}[\mathrm{osp}^{\left(2\right)}\left(2|4\right)]$
vertex-models were also obtained (see appendix). In the future, we
intend to study the $K$-matrices for the multiparametric $U_{q}[\mathrm{osp}^{\left(2\right)}\left(2|2m\right)]$
vertex-model (\emph{i.e.}, the corresponding reflection $K$-matrices
for any possible value of $\kappa_{1}$, $\kappa_{2}$, and $\nu$)
as well as the reflection $K$-matrices associated to the most general
$U_{q}[\mathrm{osp}^{\left(2\right)}\left(2n+2|2m\right)]=U_{q}[D^{\left(2\right)}\left(n+1|m\right)]$
vertex-model. 

We believe that this work contributes significantly to the classification
of the reflection $K$-matrices associated to quantum twisted Lie
superalgebras.
\begin{acknowledgements}
We thank to the referees for their valuable suggestions. The work
of Vieira has been supported by São Paulo Research Foundation (FAPESP),
grant \#2012/02144-7 and \#2011/18729-1. The work of Lima-Santos was
supported by the Brazilian Research Council (CNPq), grant \#310625/2013-0
and FAPESP, grant \#2011/18729-1.
\end{acknowledgements}

\appendix 

\section*{Appendix}

\section{Special solutions}

It is a very know fact that low-dimensional Lie (super)algebras present
special properties, for instance, being isomorphic to other Lie (super)algebras.
This also happens with the low-dimensional cases of the $\mathrm{osp}^{\left(2\right)}\left(2|2m\right)=C^{\left(2\right)}\left(m+1\right)$
Lie superalgebras considered here. In fact, it can be shown that the
$\mathrm{osp}^{\left(2\right)}\left(2|2m\right)=C^{\left(2\right)}(2)$
Lie superalgebra is isomorphic to $\mathrm{sl}^{\left(2\right)}\left(2|1\right)=A^{\left(2\right)}\left(1|0\right)$
Lie superalgebra, as well as the $\mathrm{osp}^{\left(2\right)}\left(2|4\right)=D^{\left(2\right)}(1|2)$
Lie superalgebra is isomorphic to $D^{\left(2\right)}\left(2,1,1\right)$
Lie superalgebra and, finally, that there is no other isomorphisms
for the higher values of $m$ (except those associated with an exchange
of the even and odd part, of course) \cite{FeingoldFrenkel1985,Frappat1989,Frappat2000,NeebPianzola2010,Serganova2011,Musson2012,Ransingh2013,Sthanumoorthy2016,Xu2016}. 

The existence of these special isomorphisms for the low-dimensional
Lie superalgebras $\mathrm{osp}^{\left(2\right)}\left(2|2\right)$
and $\mathrm{osp}^{\left(2\right)}\left(2|4\right)$ lead to additional
reflection $K$-matrices for the $U_{q}[\mathrm{osp}^{\left(2\right)}\left(2|2\right)]$
and $U_{q}[\mathrm{osp}^{\left(2\right)}\left(2|4\right)]$ vertex-models.
The existence of these \emph{special solutions} can be noticed directly
from the form of the complete solution presented in the section \ref{subsec:The-main-solution}.
Indeed, we can see that the complete reflection $K$-matrix of the
$U_{q}[\mathrm{osp}^{\left(2\right)}\left(2|2m\right)]=U_{q}[C^{\left(2\right)}\left(m+1\right)]$
vertex-model contain $m$ boundary free-parameters, namely, $\beta_{1,m+2}$,
$\beta_{1,m+3}$, $\ldots,$ $\beta_{1,N-2}$ and $\beta_{1,N-1}$
and, among these parameters, only $\beta_{1,m+2}$, $\beta_{1,m+3}$
and $\beta_{1,N-1}$ appear explicitly in the solution. However, we
can notice that for the cases $m=1$ or $m=2$ (but not for higher
values of $m$) some of these free-parameters become coincident. For
instance, we have $\beta_{1,m+2}=\beta_{1,N-1}$ for $m=1$ and $\beta_{1,m+3}=\beta_{1,N-1}$
for $m=2$. This fact suggests the complete solution derived in the
section \ref{subsec:The-main-solution} may not represent the most
general solution for the $U_{q}[\mathrm{osp}^{\left(2\right)}\left(2|2\right)]$
and $U_{q}[\mathrm{osp}^{\left(2\right)}\left(2|4\right)]$ vertex-models
and indeed this is the case. In fact, solving the boundary YB equation
for these two models separately, we found that are other new solutions
which hold only for these specific models (the complete solution presented
at section \ref{subsec:The-main-solution} still holds, but there
are other additional solutions that holds \emph{only} to these cases).
These special solutions will be presented in the sequel. 

\subsection{Special solutions for $U_{q}[\mathrm{osp}^{\left(2\right)}\left(2|2\right)]=U_{q}[C^{\left(2\right)}\left(2\right)]$
vertex-model}

For the $U_{q}[\mathrm{osp}^{\left(2\right)}\left(2|2\right)]=U_{q}[C^{\left(2\right)}\left(2\right)]$
vertex-model, the corresponding $K$-matrix is a four-by-four$R$-matrix:
\begin{equation}
K\left(x\right)=\left[\begin{array}{cccc}
k_{1,1}\left(x\right) & k_{1,2}\left(x\right) & k_{1,3}\left(x\right) & k_{1,4}\left(x\right)\\
k_{2,1}\left(x\right) & k_{2,2}\left(x\right) & k_{2,3}\left(x\right) & k_{2,4}\left(x\right)\\
k_{3,1}\left(x\right) & k_{3,2}\left(x\right) & k_{3,3}\left(x\right) & k_{3,4}\left(x\right)\\
k_{4,1}\left(x\right) & k_{4,2}\left(x\right) & k_{4,3}\left(x\right) & k_{4,4}\left(x\right)
\end{array}\right].
\end{equation}
The boundary YB equation consists in this case to a system of sixteen
functional equations for the elements $k_{a,b}(x)$, $1\leq a,b\leq4$.
By solving directly these equations, we found that there is only one
particular solution which is characterized by $m+2=3$ boundary free-parameters. 

The solution is the following: for the elements of the $K$-matrix
lying on the first line, we have
\begin{equation}
k_{1,2}\left(x\right)=\left(\frac{\beta_{+}+x\beta_{-}}{\beta_{1,4}}\right)G_{1}(x)k_{1,4}(x),
\end{equation}
\begin{equation}
k_{1,3}\left(x\right)=\left(\frac{\beta_{+}-\beta_{-}}{\beta_{1,4}}\right)G_{1}(x)k_{1,4}(x),
\end{equation}
 and, for the elements in the first column,
\begin{equation}
k_{2,4}\left(x\right)=ix\sqrt{q}\left(\frac{x\beta_{+}-q^{-1}\beta_{-}}{\beta_{1,4}}\right)G_{1}(x)k_{1,4}(x),
\end{equation}
\begin{equation}
k_{3,4}\left(x\right)=ix\sqrt{q}\left(\frac{x\beta_{+}+q^{-1}\beta_{-}}{\beta_{1,4}}\right)G_{1}(x)k_{1,4}(x).
\end{equation}
For the elements in the last line, we have 
\begin{equation}
k_{2,1}\left(x\right)=2\left[\frac{i\sqrt{q}}{\left(q+1\right)}\beta_{1,4}-\left(\frac{q\beta_{+}^{2}-\beta_{-}^{2}}{q-1}\right)\right]\left(\frac{\beta_{+}+x\beta_{-}}{\beta_{1,4}}\right)\frac{G_{1}(x)k_{1,4}(x)}{\beta_{1,4}^{2}},
\end{equation}
\begin{equation}
k_{3,1}\left(x\right)=2\left[\frac{i\sqrt{q}}{\left(q+1\right)}\beta_{1,4}-\left(\frac{q\beta_{+}^{2}-\beta_{-}^{2}}{q-1}\right)\right]\left(\frac{\beta_{+}-x\beta_{-}}{\beta_{1,4}}\right)\frac{G_{1}(x)k_{1,4}(x)}{\beta_{1,4}^{2}},
\end{equation}
 and, for that on the last column, 
\begin{equation}
k_{4,2}\left(x\right)=2ix\sqrt{q}\left[\frac{i\sqrt{q}}{\left(q+1\right)}\beta_{1,4}-\left(\frac{q\beta_{+}^{2}-\beta_{-}^{2}}{q-1}\right)\right]\left(\frac{x\beta_{+}+q^{-1}\beta_{-}}{\beta_{1,4}}\right)\frac{G_{1}(x)k_{1,4}(x)}{\beta_{1,4}^{2}},
\end{equation}
\begin{equation}
k_{4,3}\left(x\right)=2ix\sqrt{q}\left[\frac{i\sqrt{q}}{\left(q+1\right)}\beta_{1,4}-\left(\frac{q\beta_{+}^{2}-\beta_{-}^{2}}{q-1}\right)\right]\left(\frac{x\beta_{+}+q^{-1}\beta_{-}}{\beta_{1,4}}\right)\frac{G_{1}(x)k_{1,4}(x)}{\beta_{1,4}^{2}}.
\end{equation}
 Notice that in the present case, we have,
\begin{equation}
\beta_{\pm}=\frac{1}{2}\left(\beta_{1,2}\pm\beta_{1,3}\right),\qquad\text{and}\qquad G_{1}(x)=\frac{2}{\left(x^{2}+1\right)}.
\end{equation}

Besides, for the elements lying on the secondary diagonal, not in
the center of the $K$-matrix, we have, 
\begin{equation}
k_{1,4}\left(x\right)=\frac{1}{2}\left(x^{2}-1\right)\beta_{1,4},
\end{equation}
 and 
\begin{equation}
k_{4,1}\left(x\right)=4\left[\frac{i\sqrt{q}}{q+1}\beta_{14}-\left(\frac{q\beta_{+}^{2}-\beta_{-}^{2}}{q-1}\right)\right]^{2}\frac{k_{1,4}(x)}{\beta_{1,4}^{4}}.
\end{equation}

For the elements on the main diagonal, not in the center, we have,
respectively
\begin{equation}
k_{1,1}(x)=1+i\left[\left(\frac{x^{2}-1}{x^{2}+1}\right)\left(\frac{q\beta_{+}^{2}-\beta_{-}^{2}}{\sqrt{q}}\right)+2\sqrt{q}\left(\frac{\beta_{+}^{2}+\beta_{-}^{2}}{q-1}\right)\right]\frac{G_{1}\left(x\right)k_{1,4}\left(x\right)}{\beta_{1,4}^{2}},
\end{equation}
 and 
\begin{equation}
k_{4,4}(x)=x^{2}-\frac{ix^{2}}{\sqrt{q}}\left[\left(q\beta_{+}^{2}-\beta_{-}^{2}\right)+2\left(\frac{x^{2}q+1}{x^{2}+1}\right)\left(\frac{q\beta_{+}^{2}+\beta_{-}^{2}}{q-1}\right)\right]\frac{G_{1}\left(x\right)k_{1,4}\left(x\right)}{\beta_{1,4}^{2}},
\end{equation}
while those elements in the center are given by 
\begin{equation}
k_{2,2}\left(x\right)=\frac{i}{2}\left[\frac{\left(x^{2}q+1\right)\left(q\beta_{+}^{2}-\beta_{-}^{2}\right)-4qx\beta_{+}\beta_{-}}{\sqrt{q}\left(q-1\right)}\right]\frac{G_{1}\left(x\right)k_{1,4}\left(x\right)}{\beta_{1,4}},
\end{equation}
\begin{equation}
k_{3,3}\left(x\right)=\frac{i}{2}\left[\frac{\left(x^{2}q+1\right)\left(q\beta_{+}^{2}-\beta_{-}^{2}\right)+4qx\beta_{+}\beta_{-}}{\sqrt{q}\left(q-1\right)}\right]\frac{G_{1}\left(x\right)k_{1,4}\left(x\right)}{\beta_{1,4}}.
\end{equation}
and 
\begin{equation}
k_{3,2}(x)=k_{2,3}\left(x\right)=ix^{2}\sqrt{q}\left(\frac{\beta_{+}^{2}+q^{-1}\beta_{-}^{2}}{\beta_{1,4}^{2}}\right)G_{1}\left(x\right)^{2}k_{1,4}(x).
\end{equation}

At this point all elements of the $K$-matrix were eliminated and
we get a solution with three free-parameters, namely, $\beta_{1,2}$,
$\beta_{1,3}$ and $\beta_{1,4}$.

Finally, we remark that the $U_{q}[\mathrm{osp}^{\left(2\right)}\left(2|2\right)]=U_{q}[C^{\left(2\right)}\left(m+1\right)]$
vertex-model considered in this appendix is not related to the Yang-Zhang
vertex-model introduced in \cite{YangZhang1999} (see also \cite{GouldEtal1997,YangZhang1999,GouldZhang2000,KhoroshkinLukierskiTolstoi2001,MackayZhao2001,YangZhen2001}),
although the symmetry behind both models is the same. In fact, we
considered here the $R$-matrix introduced by Galleas and Martins
in \cite{Galleas2006} which (for $n=0$, $m=1$) correspond to a
\emph{four-dimensional} representation of the $U_{q}[\mathrm{osp}^{\left(2\right)}\left(2|2\right)]=U_{q}[C^{\left(2\right)}\left(m+1\right)]$
quantum twisted Lie superalgebra, which leads to a \emph{thirty-six
vertex-model}. On the other hand, the Yang-Zhang vertex-model \cite{YangZhang1999}
is constructed from a \emph{three-dimensional} representation of the
$U_{q}[\mathrm{osp}^{\left(2\right)}\left(2|2\right)]=U_{q}[C^{\left(2\right)}\left(m+1\right)]$
quantum twisted Lie superalgebra, which leads to a \emph{nineteen
vertex-model}. The reflection $K$-matrices of the Yang-Zhang vertex-model
were recently presented by us in \cite{VieiraLima2017A} and its algebraic
Bethe Ansatz was performed in \cite{VieiraLima2017B}.

\subsection{Special solutions for $U_{q}[\mathrm{osp}^{\left(2\right)}\left(2|4\right)]=U_{q}[C^{\left(2\right)}\left(3\right)]$
vertex-model}

For the $U_{q}[\mathrm{osp}^{\left(2\right)}\left(2|4\right)]=U_{q}[C^{\left(2\right)}\left(3\right)]$
vertex-model, the $K$-matrix is a six-by-six matrix. Besides the
complete solution presented at section \ref{subsec:The-main-solution},
there is a special solution which holds only for $m=2$ that has a
shape which resembles a \emph{X}-block matrix: 
\begin{equation}
K\left(x\right)=\left[\begin{array}{cccccc}
k_{1,1}\left(x\right) & k_{1,2}\left(x\right) & 0 & 0 & k_{1,5}\left(x\right) & k_{1,6}\left(x\right)\\
k_{2,1}\left(x\right) & k_{2,2}\left(x\right) & 0 & 0 & k_{2,5}\left(x\right) & k_{2,6}\left(x\right)\\
0 & 0 & k_{3,3}\left(x\right) & 0 & 0 & 0\\
0 & 0 & 0 & k_{4,4}\left(x\right) & 0 & 0\\
k_{5,1}\left(x\right) & k_{5,2}\left(x\right) & 0 & 0 & k_{5,5}\left(x\right) & k_{5,6}\left(x\right)\\
k_{6,1}\left(x\right) & k_{6,2}\left(x\right) & 0 & 0 & k_{6,5}\left(x\right) & k_{6,6}\left(x\right)
\end{array}\right].
\end{equation}
The elements of the $K$-matrix are the following: for the non-diagonal
elements, we have, 
\begin{align}
k_{1,2}(x) & =\left(\frac{\beta_{1,2}}{\beta_{1,6}}\right)G_{2}\left(x\right)k_{1,6}\left(x\right), & k_{1,5}\left(x\right) & =\left(\frac{\beta_{1,5}}{\beta_{1,6}}\right)G_{2}\left(x\right)k_{1,6}\left(x\right),\\
k_{2,1}(x) & =\left(\frac{\beta_{2,1}}{\beta_{1,6}}\right)G_{2}\left(x\right)k_{1,6}\left(x\right), & k_{5,1}(x) & =\left(\frac{\beta_{5,1}}{\beta_{1,6}}\right)G_{2}\left(x\right)k_{1,6}\left(x\right),\\
k_{6,2}(x) & =-x^{2}q\left(\frac{\beta_{5,1}}{\beta_{1,6}}\right)G_{2}\left(x\right)k_{1,6}\left(x\right), & k_{6,5}(x) & =-x^{2}\left(\frac{\beta_{2,1}}{\beta_{1,6}}\right)G_{2}\left(x\right)k_{1,6}\left(x\right),\\
k_{2,6}(x) & =x^{2}q\left(\frac{\beta_{1,5}}{\beta_{1,6}}\right)G_{2}\left(x\right)k_{1,6}\left(x\right), & k_{5,6}(x) & =-x^{2}\left(\frac{\beta_{1,2}}{\beta_{1,6}}\right)G_{2}\left(x\right)k_{1,6}\left(x\right),
\end{align}
with 
\begin{equation}
\beta_{2,1}=q\left(\frac{\beta_{1,5}}{\beta_{1,6}}\right)\left(\frac{2}{q+1}-\frac{\beta_{1,2}\beta_{1,5}}{\beta_{1,6}}\right),\qquad\beta_{5,1}=-\left(\frac{\beta_{1,2}}{\beta_{1,6}}\right)\left(\frac{2}{q+1}-\frac{\beta_{1,2}\beta_{1,5}}{\beta_{1,6}}\right).
\end{equation}
Notice that, in this case, 
\begin{equation}
G_{2}(x)=\frac{q+1}{qx^{2}+1}.
\end{equation}
 The elements on the secondary diagonal are given by 
\begin{align}
k_{1,6}(x) & =\frac{1}{2}\left(x^{2}-1\right)\beta_{1,6}, & k_{6,1}(x) & =-\frac{1}{q}\left(\frac{\beta_{2,1}}{\beta_{1,5}}\right)^{2}k_{1,6}(x),\\
k_{2,5}(x) & =-\left(\frac{\beta_{2,1}}{\beta_{1,2}}\right)k_{1,6}(x), & k_{5,2}(x) & =-\left(\frac{\beta_{5,1}}{\beta_{1,5}}\right)k_{1,6}(x).
\end{align}
and the elements on the main diagonal are,
\begin{align}
k_{1,1}(x) & =1-q\left(\frac{\beta_{1,2}\beta_{1,5}}{\beta_{1,6}^{2}}\right)G_{2}(x)k_{1,6}(x), & k_{5,5}\left(x\right) & =x^{2}k_{1,1}\left(x\right),\\
k_{2,2}(x) & =1+\left(\frac{\beta_{1,2}\beta_{1,5}}{\beta_{1,6}^{2}}\right)G_{2}(x)k_{1,6}(x), & k_{6,6}\left(x\right) & =x^{2}k_{2,2}\left(x\right).
\end{align}
 Finally, for the central elements, we have,
\begin{equation}
k_{4,4}(x)=k_{3,3}\left(x\right)=\left[\frac{1}{G\left(x\right)}-\left(\frac{x^{2}q^{2}-1}{q+1}\right)\left(\frac{\beta_{1,2}\beta_{1,5}}{\beta_{1,6}^{2}}\right)G(x)k_{1,6}\left(x\right)\right].
\end{equation}

With this the boundary YB equation is completely satisfied. We get
as well a solution with $3$ boundary free-parameters, namely, $\beta_{1,2}$,
$\beta_{1,5}$ and $\beta_{1,6}$.

We also report existence of the special diagonal solution $K\left(x\right)=\mathrm{diag}\left(1/x^{2},1,1,1,1,x^{2}\right)$,
which holds both for the $U_{q}[\mathrm{osp}^{\left(2\right)}\left(2|4\right)]=U_{q}[C^{\left(2\right)}\left(3\right)]$
and $U_{q}[\mathrm{osp}^{\left(2\right)}\left(6|0\right)]=U_{q}[D_{3}^{(2)}]$
vertex-models \cite{Guan2000,Lima2001,Vieira2013}.

\bibliographystyle{iopart-num}
\bibliography{references}

\end{document}